\documentclass[aps,prl,amsmath,amssymb,superscriptaddress,showkeys,showpacs,twocolumn,floatfix]{revtex4-1}

\usepackage[final]{graphicx}
\usepackage{hyperref}
\usepackage{amsmath}
\usepackage{bbm}
\usepackage{amsfonts}
\usepackage{amssymb}
\usepackage{latexsym}
\usepackage{graphicx}
\usepackage[english]{babel}
\usepackage{multirow}
\usepackage{float}
\usepackage{url}
\usepackage{slashed}
\usepackage{xcolor}
\usepackage[utf8]{inputenc}
\usepackage{verbatim}
\usepackage{lipsum}

\usepackage[utf8]{inputenc}

\usepackage{diagbox}

\usepackage{mathtools}
\usepackage{amsfonts}
\usepackage{mathrsfs}
\usepackage{bbm}
\usepackage{slashed}

\usepackage{graphicx}
\usepackage{color}
\usepackage{array}
\usepackage{esint}
\usepackage{placeins}
\usepackage{booktabs}
\usepackage{makecell}
\usepackage{epstopdf}
\usepackage[caption=false]{subfig}

\usepackage{xspace}
\usepackage{siunitx}
\usepackage{hyperref}
\usepackage[nameinlink]{cleveref}
\usepackage{appendix}

\usepackage{xifthen}
\usepackage{xcolor}
\hypersetup{
	colorlinks,
	linkcolor={red!75!black},
	citecolor={blue!75!black},
	urlcolor={blue!75!black}
}
\usepackage{comment}

\setkeys{Gin}{width=0.48\textwidth}
\graphicspath{
	{./figures/}
	{../figures/}
}

\def\s0#1#2{\mbox{\small{$ \frac{#1}{#2} $}}}
\def\0#1#2{\frac{#1}{#2}}

\usepackage{xcolor}

\usepackage{ulem}

\begin{document}
	\title{First-order Electroweak phase transition with Gauge-invariant approach}

\author{Renhui Qin}
\affiliation{Department of Physics and Chongqing Key Laboratory for Strongly Coupled Physics, Chongqing University, Chongqing 401331, P. R. China}

\author{Ligong Bian}
\email{lgbycl@cqu.edu.cn}
\affiliation{Department of Physics and Chongqing Key Laboratory for Strongly Coupled Physics, Chongqing University, Chongqing 401331, China.}
\affiliation{Center for High Energy Physics, Peking University, Beijing 100871, China.}

\date{\today}
	
\begin{abstract}
We study the electroweak phase transition dynamics with a three-dimensional standard model effective field theory under a gauge-invariant approach.  We observe that, at the two-loop level, the phase transition parameters obtained with the gauge invariant approach can at most deviate from that of the dimensional reduction method around the percent level. We further found that the predicted gravitational wave signals at the new physics scale $\Lambda\gtrsim 590$ GeV are unreachable by the space-based interferometers, such as: LISA, TianQin, and Taiji.
 
\end{abstract}
\maketitle
	
\subparagraph{Introduction}  
 First-order phase transition (PT) is an important way to study the evolution of the universe, which provides a chance to generate baryon asymmetry of the Universe through electroweak baryogenesis~\cite{Morrissey:2012db}, primordial magnetic field~\cite{Yang:2021uid,Di:2020kbw,Zhang:2019vsb,Stevens:2012zz,Kahniashvili:2009qi,Hindmarsh:1997tj,Grasso:1997nx,Ahonen:1997wh}, and one important stochastic gravitational wave (GW) source to be probed by the next-generation GW detector experiments, such as LISA \cite{Athron:2023xlk,Caprini:2015zlo}. Such a PT can be achieved in many models beyond the Standard Model (SM) since the SM of particle physics with the $125$ GeV Higgs admits a smooth crossover between the high and low temperature phases~\cite{DOnofrio:2014rug}.  

 The dynamical symmetry breaking process driven by the first-order PT can be quantitively calculated by the thermal effective potential~\cite{Athron:2023xlk}. The studies of the PT dynamics are usually performed on the effective potential in the Landau gauge and therefore suffer from the gauge dependent problem due to the element fields are not invariant under gauge transformation~\cite{Hu:1996qa,Patel:2011th}. However, since the physical observables must be gauge independent, the gauge invariant (GI) results from gauge dependent theory should be achieved. 
The Nielsen identity shows that though the ordinary effective potential is gauge dependent, its minimal value is gauge independent~\cite{Hu:1996qa,Nielsen:1975fs}. One way to maintain the GI is to perform $\hbar-$expansion in the PT dynamics calculation, where the Nielsen identities guarantee the GI order-by-order, see studies in the abelian Higgs model~\cite{Garny:2012cg,Hirvonen:2021zej,Lofgren:2021ogg} and the SM extend by scalar model~\cite{Athron:2022jyi} in the 4D thermal field theory. Refs.~\cite{Schicho:2022wty,Niemi:2020hto,Croon:2020cgk} further apply the method to the 3D thermal field theory after performing dimensional reduction (DR). The DR can build a theory at a light scale after integrating out superheavy and heavy modes.  Wherein, the temperature only enters the parameters of the light theory, and there is no IR problem due to these fields are massive. 

Another alternative way to achieve the GI results is to construct a new potential at the minimum of the effective potential. This approach is to define the external source coupling to a gauge invariant composite operator $\mathcal{L}\rightarrow \mathcal{L}-J\langle\Phi^\dagger\Phi\rangle$~\cite{Hu:1996qa,Buchmuller:1994vy,Hebecker:1995kd,Buchmuller:1995sf}. Therefore, such an effective potential can be obtained after performing the Legendre transform of the ground state energy density. We construct the GI potential at three dimensions thermal field theory after conducting the procedure of the DR~\cite{Kajantie:1995dw,Farakos:1994kx,Farakos:1994xh,Laine:1994zq}. 

In this Letter, with SM effective field theory (SMEFT) as a concrete illustration,  we construct the GI thermal effective potential based on the 3D effective potential at 2-loop order
after considering that the composite operator $\langle\Phi^\dagger\Phi\rangle$ coupled to the external source. 
We assess the reliability of the 
GI composite thermal effective potential by quantitively computing the PT dynamics. It has been established that the 3D thermal SMEFT at 2-loop level will reduce the renormalization scale dependence fairly well~\cite{Qin:2024idc,Croon:2020cgk}\footnote{For the research with 4d SMEFT, see~\cite{Cai:2017tmh,Grojean:2004xa,Delaunay:2007wb,Chala:2018ari,Bodeker:2004ws,Hashino:2022ghd,Damgaard:2015con,Postma:2020toi}}. 
We estimate the difference of PT quantities between the DR and GI approaches and show that the strong first-order electroweak PT occurs in the new physics (NP) scale range of 590 GeV $\lesssim\Lambda\lesssim 800$ GeV. In addition, we investigate the implications of the GW predictions. 

\subparagraph{Dimensional redution.}
 Since the potential barrier for SMEFT mostly comes from the dimension-six operator of $(\Phi^\dag\Phi)^3/\Lambda$ with $\Lambda$ being the NP scale, we consider the SMEFT with that single dimension-six operator in this Letter. The original 4d Lagrangian has the form
\begin{align}
\mathcal{L}=&-\frac{1}{4}G_{\mu\nu}^a G_{\mu\nu}^a-\frac{1}{4}F_{\mu\nu}F^{\mu\nu}+(D_\mu\Phi)^\dag(D_\mu\Phi)\nonumber\\
&+\mu_h^2\Phi^\dag \Phi+\lambda (\Phi^\dag \Phi)^2+c_6 (\Phi^\dag \Phi)^3\;,
\end{align}
where $c_6=\Lambda^{-2}$ and $\Phi$ is the SM Higgs doublet. This model in three dimensions is described by the Lagrangian
\begin{align}\label{lightmode}
\mathcal{L}=\frac{1}{4}G_{ij}^{a}G_{ij}^{a}
    + &\frac{1}{4}F_{ij}^{ }F_{ij}^{ }
    + (D_{i}\Phi^{ })^\dagger (D_{i}\Phi^{ })
    + V
    \;,
\end{align}
where
$G^{a}_{ij} =
\partial_{i} A^a_j -
\partial_{j} A^a_i + g_{3} \epsilon^{abc}A_{i}^b A_{j}^c$,
$F_{ij} = \partial_{i}B_{j} - \partial_{j}B_{i}$ and
$D_{i}\Phi = (\partial_{i} - ig_{3} \tau^a A_{i}^a/2 - ig^\prime_{3} B^{ }_{i}/2)\Phi^{ }$ with $\tau^a$ being the Pauli matrices. The fields in Eq.~\eqref{lightmode} with dimension $\text{GeV}^{1/2}$ and parameters are dimensionful. This theory works when the gauge couplings fulfill the condition of $g<1$. The DR theory spilled the mass scale with superheavy ($\pi T$), heavy ($g T$), and light mode ($g^2 T$). The theory is simpler than the 4d theory in that the fermions and bosons with masses being larger than $gT$ have been integrated out. 
The $V$ is the light mode potential which was obtained at the 2-loop order after integrating out superheavy and heavy mode at Landau gauge and has the form
\begin{equation}\label{veff}
V=V_{\rm tree}+\hbar V_{\rm 1loop}+\hbar^2 V_{\rm 2loop}\;,
\end{equation}
where
\begin{align}\label{tree}
V_{\rm tree}=&
    \frac{1}{2} \mu_h^2 \phi^2
    + \frac{1}{4} \lambda \phi^4
    + \frac{1}{8} c_6 \phi^6 \;, \\
V_{\rm 1loop}=&
(d-1)\Big( 2 J_{soft}(m_{W}) + J_{soft}(m_{Z}) \Big)\nonumber\\
    &+J_{soft}(m_{\phi}) + 3 J_{soft}(m_{\chi}) \;,
\end{align}
with
\begin{equation}
 J_{soft}(m) =
- \frac{1}{12\pi} (m^2)^\frac{3}{2}
\;.
\end{equation}
Therein, the field-dependent masses  are
\begin{align}
m^2_{\phi}&=\mu_{h}^2+ 3\lambda\phi^2+ \frac{15}{4} c_{6} \phi^4
\;,
m^2_{W}=\frac{1}{4}g^2 \phi^2
\;,\\
m^2_{\chi}&=\mu_{h}^2
    +\lambda  \phi^2
    + \frac{3}{4} c_{6} \phi^4
\;, m^2_{Z,3}=\frac{1}{4} \Big( g^2+ g^{\prime 2} \Big) \phi^2\;.
\end{align}

The $2$-loop effective potential $V_{\rm 2loop}$ and related parameters can be found in Ref.~\cite{Qin:2024idc}.

\subparagraph{Gauge invariant potential.}
   The GI potential can be constructed by introduce a current $J$ coupling to composite field $\sigma=\phi^\dag \phi$\cite{Buchmuller:1994vy,Hu:1996qa,Hebecker:1995kd,Buchmuller:1995sf}
   \begin{equation}
    e^{- W[J]}=\int \mathcal{D}A_i^a\mathcal{D}\phi e^{-S-\int d^3 x J\sigma}\;,
   \end{equation}
  then the effective potential is given by the Legendre transformation
   \begin{equation}\label{legendre}
     V(\sigma)=V(v(T),\mu_h^2+J)-J\sigma\;, \sigma=\partial V(v(T),\mu_h^2+J)/\partial J\;.
   \end{equation}
   The composite effective potential $V(\sigma)$ is thus gauge independent since it is constructed to be the minimal value of $W[J]$ with respect to $J$ and $\phi$~\cite{Hu:1996qa}.
   We calculate the effective potential at $\hbar^2$ order, this require the current $J$ need to $\hbar$ order $J=J_0+\hbar J_1$\cite{Farakos:1994xh}. With the definition
   \begin{equation}
    v^2=v_0^2+\hbar v_1^2+ \hbar^2 v_2^2\;,
   \end{equation}
   the effective potential at the vacuum expectation value (vev) expands as
   \begin{equation}
   \begin{aligned}
    V(v)=&V_{\rm tree}(v_0^2+\hbar v_1^2+\hbar^2 v_2^2)+\hbar V_{\rm 1loop}(v_0^2+\hbar v_1^2)\\
    &+\hbar^2 V_{\rm 2loop}(v_0^2)\;,
    \end{aligned}
   \end{equation}
   the two solutions of the condition
    \begin{equation}
    \left.\frac{d V(\phi^2)}{d(\phi^2)}\right|_{\phi^2=v^2}=0\;,
    \end{equation}
   $\phi^2_s=0$ and $\phi^2_b=v^2$ correspond to the symmetric and broken phase, respectively. For the leading order in $\hbar$
   \begin{equation}
   v_0^2=\frac{2}{3c_6}\left(-\lambda+\sqrt{-3c_6 J_0-3 c_6 \mu_h^2+\lambda^2}\right)\;,
   \end{equation}
   Insert this in the Eq.\eqref{legendre}, we can obtain the relation
   \begin{equation}
    J_0=-\mu_h^2-2\lambda \sigma-3c_6  \sigma^2\;.
   \end{equation}
   To the order of $\hbar$, we have
   \begin{equation}
    v_1^2=-\frac{V_{1loop}^\prime(\phi^2)}{V_{tree}^\prime(\phi^2)}\;.
   \end{equation}
   By defining
   \begin{equation}
   \begin{aligned}
     m_W=&\sqrt{\frac{1}{2}g^2\sigma}, m_Z=\sqrt{\frac{1}{2}(g^2+g^{\prime 2})\sigma},\\ m_\phi=&\sqrt{4\lambda\sigma+12c_6 \sigma^2}\;,
   \end{aligned}
   \end{equation}
   the $J_1$ can be simplified to
   \begin{equation}
     J_1=\frac{1}{8\pi}\left[(4(\lambda+6 c_6 \sigma)m_\phi+2g^2m_W+(g^2+g^{\prime 2}m_Z)\right]\;.
   \end{equation}
The 2-loop GI potential in the broken phase is the
\begin{equation}
\begin{aligned}
V_b(\sigma)=&\mu_h^2\sigma+\lambda\sigma^2+c_6\sigma^3-\frac{\hbar}{12\pi}(4m_W^3+2m_Z^3+m_\phi^3)\\
     &-\frac{\hbar^2}{64\pi^2}\left[(5\lambda+27c_6\sigma)m_\phi^2+2g^2m_W m_\phi\right.\\
     &\left.+(g^2+g^{\prime 2})m_Z m_\phi\right]+\hbar^2 V_{\rm 2loop}(m_\chi=0)\;,
\end{aligned}
\end{equation}
where the $V_{\rm 2loop}(m_\chi=0)$ is a function of $\sigma,m_W,m_Z,m_\phi$ with $m_\chi=0$. For the symmtric phase $(\phi=0)$, the leading order of Eq.\eqref{veff} gives
\begin{equation}
 \frac{\partial V}{\partial J}=\hbar\sqrt{\mu_h^2+\hbar J_1}=-2\pi \sigma\;,
\end{equation}
it require $\sigma\leq 0$ at symmetric phase. Then we can obtain the
\begin{equation}
J_1=(4\pi^2\sigma^2-\mu_h^2\hbar^2)/\hbar^3\;,
\end{equation}
and the 2-loop GI potential at the symmetric phase is
\begin{equation}
\begin{aligned}
V_s(\sigma)=&\mu_h^2 \sigma+\frac{3}{16}\left(3g^2+g^{\prime 2}+8\lambda\right)\sigma^2-\frac{4\pi^2\sigma^3}{3\hbar^3}\\
&+\frac{1}{4}\left(3g^2+g^{\prime 2}\right)\sigma^2\log\left[-\frac{\overline{\mu}_3\hbar}{4\pi \sigma}\right]\;,
\end{aligned}
\end{equation}
where $\overline{\mu}_3=g^2T$ is the light mode renormalization scale and has little effect on the result. The GI potential  is
\begin{equation}\label{veffg}
V(\sigma)=\theta(\sigma)V_b(\sigma)+\theta(-\sigma)V_s(\sigma)\;.
\end{equation}
    \subparagraph{Nucleation criteria.}
As the universe cools to the critical temperature $T_c$, the broken and symmetric phase has the same free energy. The temperature $T_c$ can be obtained by solving
\begin{equation}
  V_b(\sigma_b)=V_s(\sigma_s), \left.\frac{\partial V_b(\sigma)}{\partial \sigma}\right|_{\sigma=\sigma_b}=0, \left.\frac{\partial V_s(\sigma)}{\partial \sigma}\right|_{\sigma=\sigma_s}=0\;.
\end{equation}
 When the temperature of the universe cools below $T_c$, the symmetric phase becomes unstable and will decay to the true phase. At the nucleation temperature $T_n$, the bubble nucleation rate
$\Gamma=Ae^{-S_c}$ is equal to the Hubble rate $\Gamma\sim H$, i.e.$S_c\approx 140$. The Euclidean action is
\begin{equation}
 S_c=\int \frac{\sigma^\prime}{2\sigma}+V(\sigma)d^3 x\;.
\end{equation}
The action can be obtained by solving the bounce function
\begin{equation}\label{bounce1}
 \frac{1}{2\sigma}\left(\frac{d^2\sigma}{d\rho^2}-\frac{1}{2\sigma}\left(\frac{d\sigma}{d\rho}\right)^2+\frac{2}{\rho}\frac{d\sigma}{d\rho}\right)=\frac{d V(\sigma)}{d \sigma}\;.
\end{equation}
After insert $\sigma=\phi^2/2$ in eq.\eqref{bounce1}, the bounce function becomes
\begin{equation}\label{bounce2}
\frac{d^2\phi}{d\rho^2}+\frac{2}{\rho}\frac{d\phi}{d\rho}=\frac{dV(\phi)}{d\phi}\;,
\end{equation}
and
\begin{equation}\label{vefftn}
\begin{aligned}
 V(\phi)=&\theta(\phi-\sqrt{2|\sigma_s|})V_b(\phi^2/2+\sigma_s)\\
 &+\theta(-(\phi-\sqrt{2|\sigma_s|}))V_s(\phi^2/2+\sigma_s)\;,
\end{aligned}
\end{equation}
 we use the code ``BubbleDet'' to solve Eq.\eqref{bounce2} and obtain the nucleation temperature $T_n$ \cite{Ekstedt:2023sqc}. Then, we evaluate the inverse PT duration $\beta/H=T(dS_c/dT)$ and PT strength $\alpha=T(\Delta\rho/\rho_{rad})$ at temperature $T_n$, where $\Delta \rho$ is the difference in the trace anomaly \cite{Hindmarsh:2017gnf,Espinosa:2010hh,Croon:2020cgk,Qin:2024idc}
\begin{equation}
\begin{aligned}
\Delta\rho&=-\frac{3}{4}\Delta V(\phi,T)+\frac{1}{4}T \frac{d \Delta V(\phi,T)}{d T}\;,\\
\Delta V&=V(\phi)-V_s(\sigma_s)\;,
\end{aligned}
\end{equation}
and $\rho_{rad}=(\pi^2/30)g_* T^4$ is the radiation density of the high-temperature phase with the effective number of relativistic degrees of freedom  $g_*=106.75$.

 \subparagraph{SFOEWPT condition.}  The EW sphaleron process provides the baryon number violation condition to explain the baryon asymmetry of the Universe\cite{Sakharov:1967dj,Kuzmin:1985mm,Shaposhnikov:1987tw,Morrissey:2012db}. The sphaleron process will reach equilibrium at high temperature, and it should be strongly suppressed at low temperature to avoid washout of the generated baryon number which corresponds to the requirement of a strongly-first order electroweak PT~\cite{Patel:2011th}, i.e. $\Gamma_{sph}\varpropto \exp{(-E_{sph}^{EW}/T)}\ll H(T)$ with $E_{sph}^{EW}$ being the electroweak sphaleron energy,
this requires that the sphaleron rate in the broken phase is lower than the Hubble rate.
After reorganizing the relation among the sphaleron energy, PT temperature ($T_n$), and the scalar VEV, the baryon number washout avoidance condition can be obtained to restrict the baryon asymmetry explanation parameter spaces~\cite{Zhou:2019uzq,Gan:2017mcv}:
\begin{equation}\label{PT2}
PT_{sph}\equiv\frac{E_{sph}}{T}-7\ln\frac{v}{T}+\ln\frac{T_n}{100\text{GeV}}>(35.9-42.8)\;.
\end{equation}

\begin{figure}[!htp]
    \centering
    \includegraphics[width=0.8\linewidth]{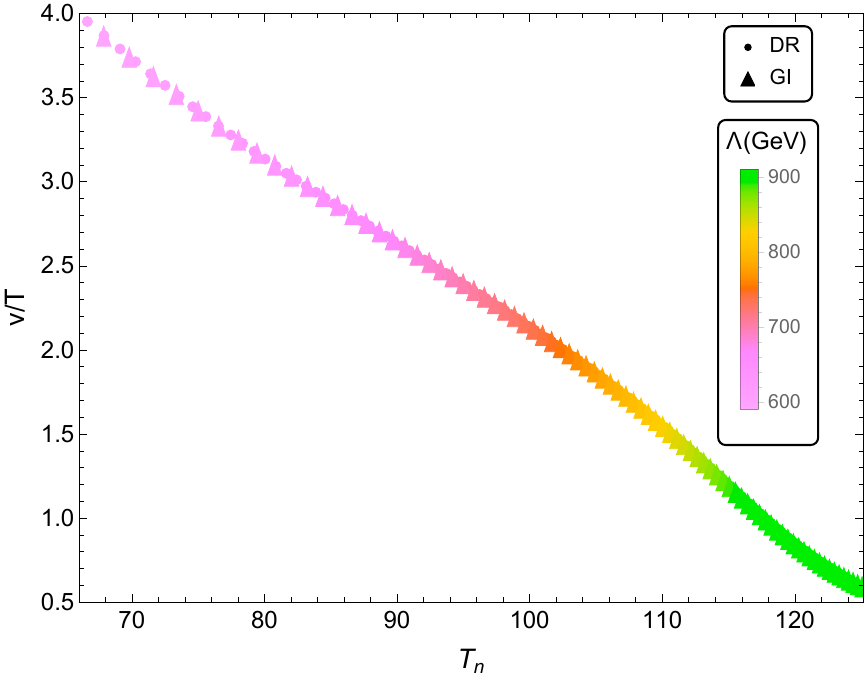}
    \caption{The nucleation temperature $T_n$ and $v/T$ as function of $\Lambda$ with DR and GI potentials. }
    \label{figtvbt}
\end{figure}

Fig.~\ref{figtvbt} shows the relation between the temperature $T_n$ and $v/T$ in DR and GI potentials at different $\Lambda$. The parameter $v/T$ is defined as $v/T=\sqrt{2\langle\Phi^\dagger \Phi\rangle}/\sqrt{T}$ with $\langle
\Phi^\dagger\Phi\rangle=d V_{eff}/d\overline{\mu}_{h,3}^2$~\cite{Qin:2024idc}, where the $\overline{\mu}_{h,3}^2$ is the light scale mass parameter, and $\overline{\mu}_{h,3}^2=\mu_h^2$ in Eq.\eqref{tree}. As the $\Lambda$ increases, the $v/T$ decreases while $T_n$ increases. For the GI we use the relation $\sigma=\sigma_b-\sigma_s$ at $T_n$, since we use the potential of Eq.~\eqref{vefftn} to solve the bounce function of Eq.~\eqref{bounce2}. The difference of $v/T$ in the DR and GI approach is insignificant. This is because that $dV_{eff}(v,\overline{\mu}_{h,3}^2)/d\overline{\mu}_{h,3}^2=dV_{eff}(v,\overline{\mu}_{h,3}^2+J)/dJ$.

\begin{figure}[!htp]
    \centering
    \includegraphics[width=0.8\linewidth]{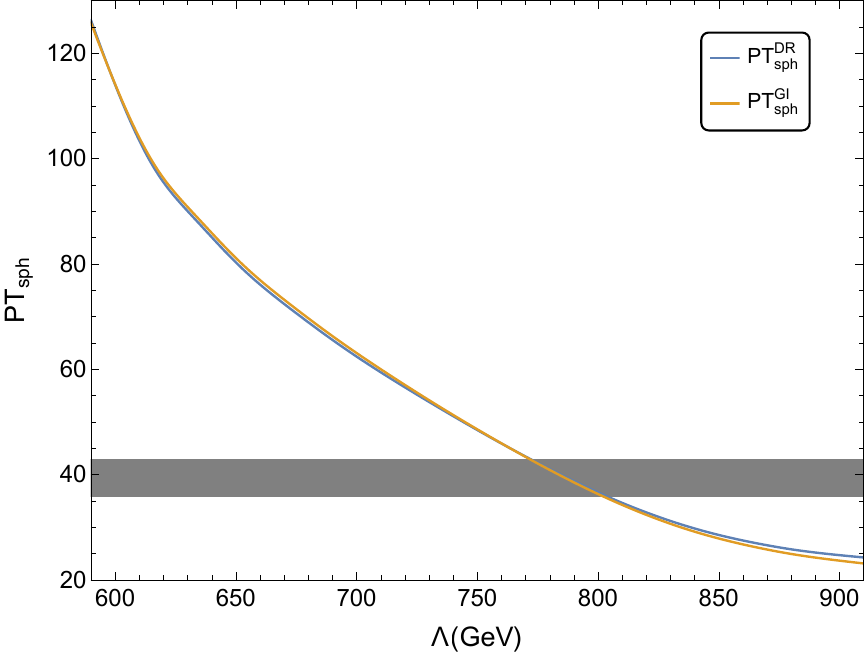}
    \caption{The parameter $PT_{sph}$ as function of $\Lambda$ with DR and GI potentials.}
    \label{figpt}
\end{figure}

We show the limitation of the model parameter $\Lambda$ by the baryon asymmetry with the baryon number violation condition \cite{Qin:2024idc,Sakharov:1967dj,Kuzmin:1985mm,Shaposhnikov:1987tw,Morrissey:2012db} in the Fig.~\ref{figpt}.  The magnitude of the parameter $PT_{\rm sph}$ decreases as the $\Lambda$ increases, and the difference between the DR and GI approach is insignificant. The DR and GI approach gives the same region of $\Lambda$, i.e., $\Lambda\lesssim 770-800$ GeV.

\begin{figure}[!htp]
    \centering
    \subfloat[The relation between the $\alpha$ and $\beta/H$.]{\includegraphics[width=0.8\linewidth]{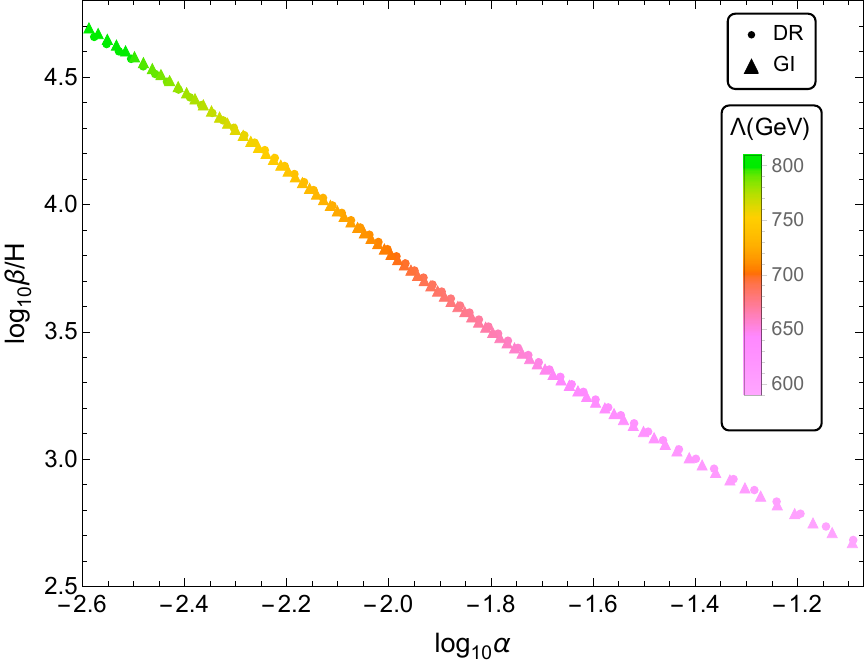}\label{alphabetah}}
    \hfill
     \subfloat[The relation between the $v_w$ and $T_n$.]{\includegraphics[width=0.8\linewidth]{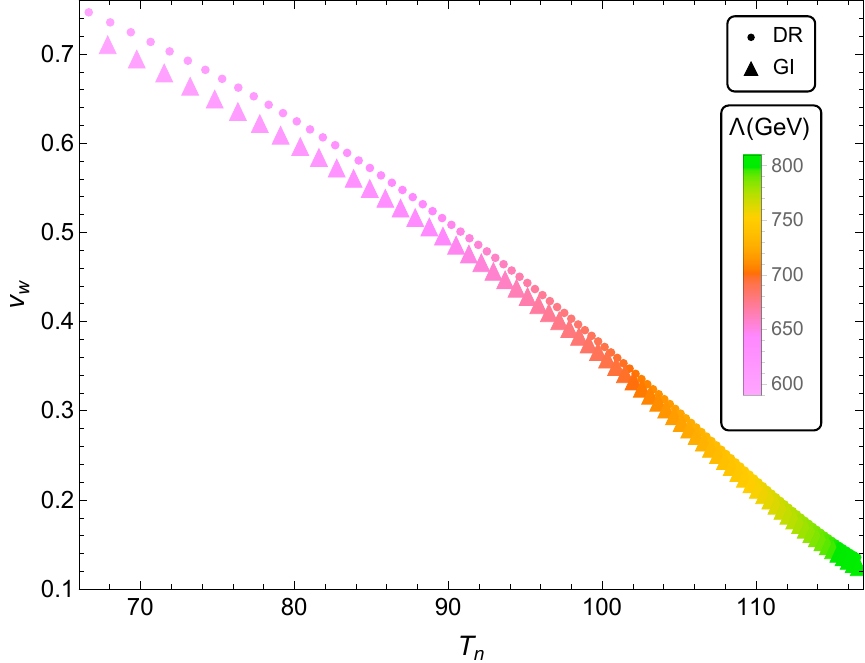}\label{vwT}}
    \caption{The PT parameters $\alpha,\beta/H,T_n$ and $v_w$ as function of $\Lambda$. The color bar shows the value of $\Lambda$.}
    \label{figalphabetahvw}
\end{figure}

\subparagraph{PT Thermodynamics.}
We show the PT parameters $\alpha,\beta/H,T_n$ and $v_w$ as function of the $\Lambda$ ($\Lambda\leq 800$ GeV) with DR and GI approaches in the Fig.~\ref{figalphabetahvw}. The top plot shows as the $\Lambda$ increases, the magnitude of the $\alpha$ ($\beta/H$) decreases (increase). The differences between the results of $\alpha$ and $\beta/H$ calculated with DR and GI are negligible. The bottom one shows that the $v_w$ ($T_n$) decreases (increases) as the NP physics scale $\Lambda$ increases. The results of $v_w$ computed with the GI approach are less than those in the DR approach, and the two gradually approach each other at large $\Lambda$ closing to $800$ GeV. This is because $v_w\propto \sqrt{\Delta V}$ (the difference between the broken phase and symmetric phase) and the $\Delta V$ in the DR approach is greater than that of the GI approach (as shown in Fig.\ref{figveff2}). Here, we noticed that, with the $\hbar-$expansion of the 3D effective potential at the minimum, Ref.~\cite{Croon:2020cgk} revealing the gauge invariant can be reached at the order of $\mathcal{O}(g^4)$. Therefore, as a byproduct, our results can serve as a double-check of that paper. 

\begin{figure}[!htp]
    \centering
 \includegraphics[width=0.8\linewidth]{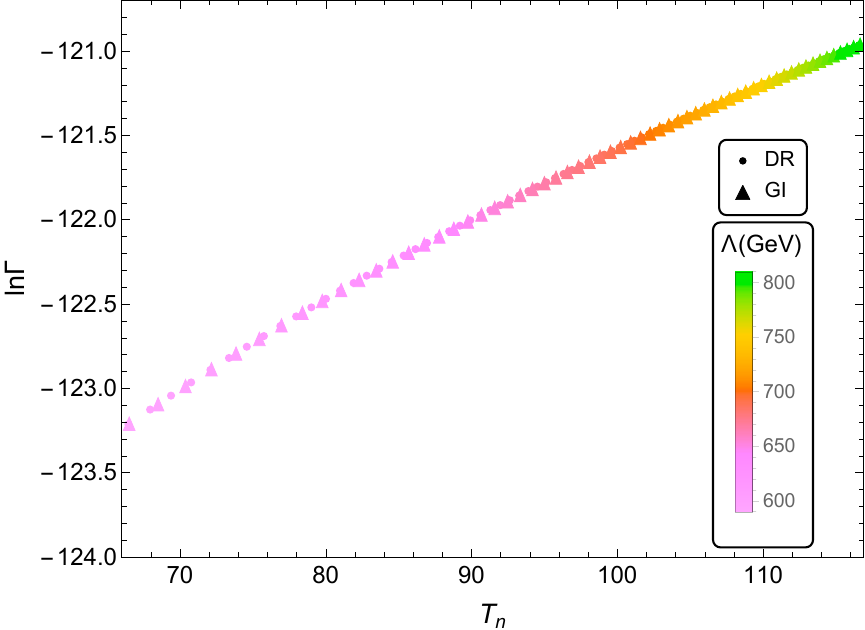}
    \caption{The nucleation rate $\Gamma$ as a function of the NP scale $\Lambda$.}
    \label{figrateA}
\end{figure}

We show that the nucleation rate $\Gamma$ increases slowly as the $\Lambda$ grows in the Fig.\ref{figrateA}. Taking the approximation $A\approx T_n^4$ and $\ln{\Gamma}=-S_c+4\ln{T_n}$,
we observe that the nucleation rate $\ln{\Gamma}\sim -120$.  We can see that the difference between the DR and GI approach is insignificant.

\begin{figure}[!htp]
    \centering
\includegraphics[width=0.8\linewidth]{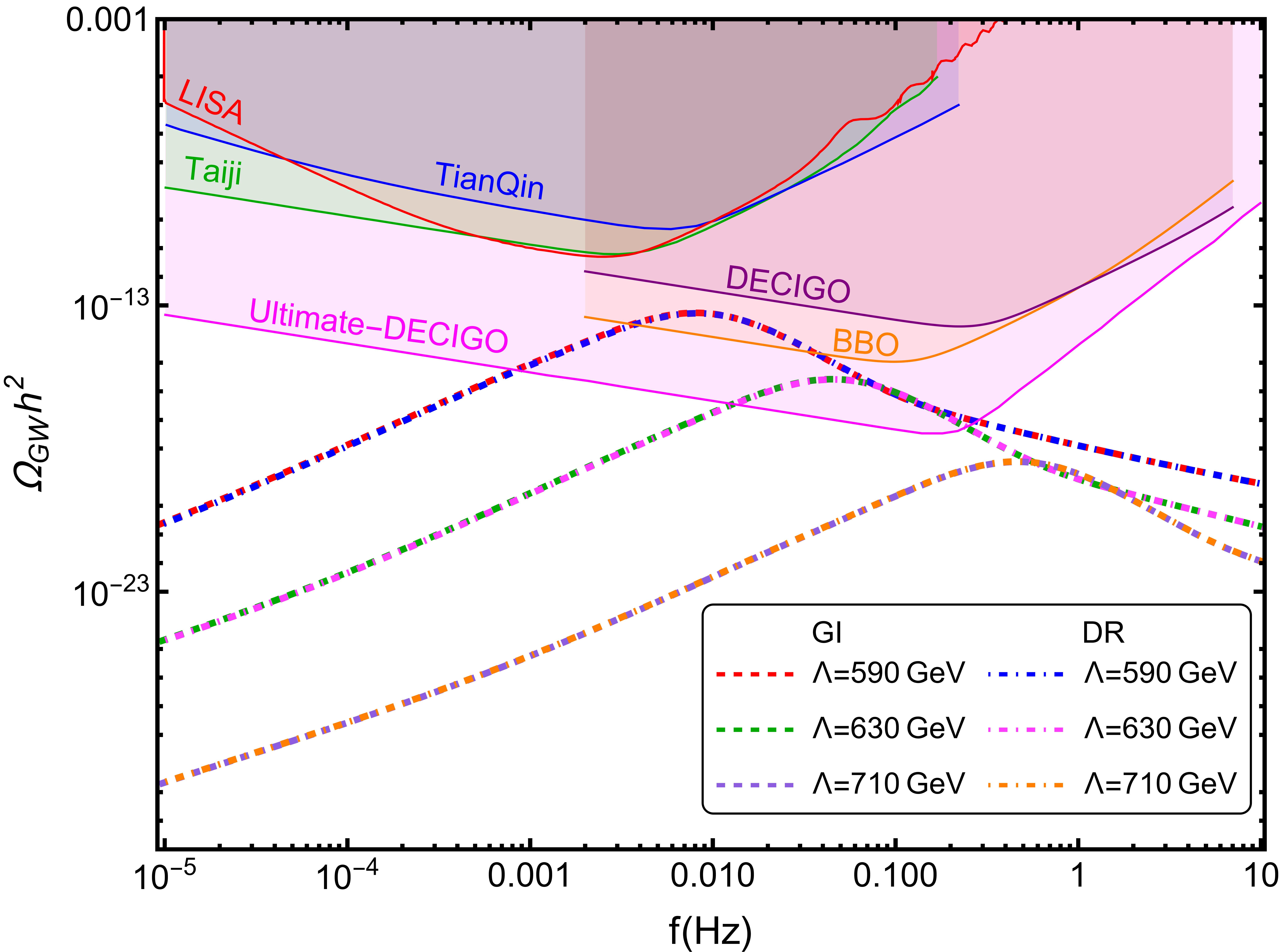}
    \caption{The GWs spectra of the first-order PTs at SMEFT. The sensitivity curves includes: Taiji~\cite{Hu:2017mde,Ruan:2018tsw}, Tianqin~\cite{TianQin:2015yph,Zhou:2023rop}, LISA~\cite{LISA:2017pwj,Baker:2019nia}, BBO~\cite{Crowder:2005nr,Corbin:2005ny,Harry:2006fi} and DECIGO~\cite{Seto:2001qf,Kawamura:2006up,Yagi:2011wg,Isoyama:2018rjb}. }
    \label{figgw}
\end{figure}

Basically, for the GWs from first-order PT, there are three sources, i.e., bubble collision~\cite{Caprini:2007xq,Huber:2008hg}, sound wave~\cite{Hindmarsh:2013xza,Hindmarsh:2015qta,Hindmarsh:2017gnf}, and MHD turbulence\cite{Caprini:2009yp}.
In the figure.\ref{figgw}, we illustrate the disparity between the computed GW spectra based on the PT parameters and the sensitivity curves of the Taiji, Tianqin, LISA, BBO, and DECIGO. The first-order phase transition at SMEFT with $\Lambda\gtrsim 590$GeV cannot be produced by the Taiji, Tianqin, and LISA. A stronger phase transition needs the NP scale to be $\Lambda\lesssim 590$ GeV, but it is could not be calculated by the DR approach since the high-temperature expansion has used to simplify the DR calculation and it would induce large uncertainties there~\cite{Croon:2020cgk}.


 \subparagraph{Conclusion.}                                                                                      In this Letter, we construct the gauge invariant thermal effective potential based on the three-dimensional thermal standard model effective field theory. We study the phase transition dynamics and find that the electroweak phase transition parameters can be strongly first-order when the new physics scale $\Lambda\lesssim 770-800$ GeV. The phase transition parameters (trace anomaly, phase transition temperature, and duration, bubble wall velocity) calculated with the gauge invariant approach and the dimensional reduction only deviate from each other at the percent level. 
 We observe that in the NP scale range of $\Lambda \in[590,910]$ GeV, the predicted GW signals in 3D thermal SMEFT 
  can be probed by BBO and Ultimate-DECIGO, but they  
 are unreachable by the space-based interferometers, such as: LISA, TianQin, and Taiji.

 We further note that one can quantitatively assess the reliability of the perturbative computations in the GI approach through comparison with the nonperturbative lattice simulation of the bubble nucleation rate in the first-order PT process~\cite{PhysRevD.63.045002,Moore:2001vf}.
 
\subparagraph{Acknowledgments}
 L.B. is supported by the National Natural Science Foundation of China (NSFC) under Grants No. 12075041, No. 12322505, and No. 12347101. L.B. also acknowledges Chongqing Talents: Exceptional Young Talents Project No. cstc2024ycjh-bgzxm0020.

\bibliography{reference}

\clearpage

\onecolumngrid
\begin{center}
  \textbf{\large Supplemental Material}\\[.2cm]
\end{center}


\section{The realtions between \texorpdfstring{$\overline{MS}$}{} parameters and physical observables}
The dimensional regularisation in the $\overline{MS}-$scheme and the renormalization group (RG) running has been taken to perform the numerical calculation of the thermal effective potential. Since within the DR scheme, the renormalization scale dependence at the 2-loop level is very slight,  we set the renormalization scale $\overline{\mu}=4\pi e^{-\gamma_E}T$, where $\gamma_E$ is the Euler-Mascheroni constant. We relate the $\overline{MS}-$parameters (gauge and Yukawa couplings) before RG running to physical observables of Fermi constant $G_F=1.1663787\times 10^{-5}\text{GeV}^{-2}$ and pole masses~\cite{ParticleDataGroup:2022pth}
\begin{equation}\label{massparameter1}
(M_t,M_W,M_Z,M_h)=(172.76,80.379,91.1876,125.1)~\text{GeV}.
\end{equation}

The tree-level Higgs mass parameter and self-interaction can be obtained by solving 
\begin{equation}
\left.\frac{\partial^2 V_{tree}}{\partial \phi^2}\right|_{\phi=v_0}=M_h^2,\quad \left.\frac{\partial V_{tree}}{\partial\phi}\right|_{\phi=v_0}=0\;,
\end{equation}
where $V_{tree}$ is defined in Eq.\eqref{tree}. Then, we can obtain
\begin{equation}
\mu_h^2=-\frac{1}{2}M_h^2+\frac{3}{4}c_6v_0^4\;,\quad  \lambda=\frac{1}{2}\frac{M_h^2}{v_0^2}-\frac{3}{2}c_6 v_0^2\; .
\end{equation}

Since the 2-loop order DR approach is reaching at $\mathcal{O}(g^4)$, the tree-level($\mathcal{O}(g^2)$) relations need to be improved by their one-loop corrections. The relevant calculations have been done, see details in Refs.\cite{Qin:2024idc,Croon:2020cgk}.


\section{Effective potential}
\begin{figure}[!htp]
    \centering
    \includegraphics[width=0.42\linewidth]{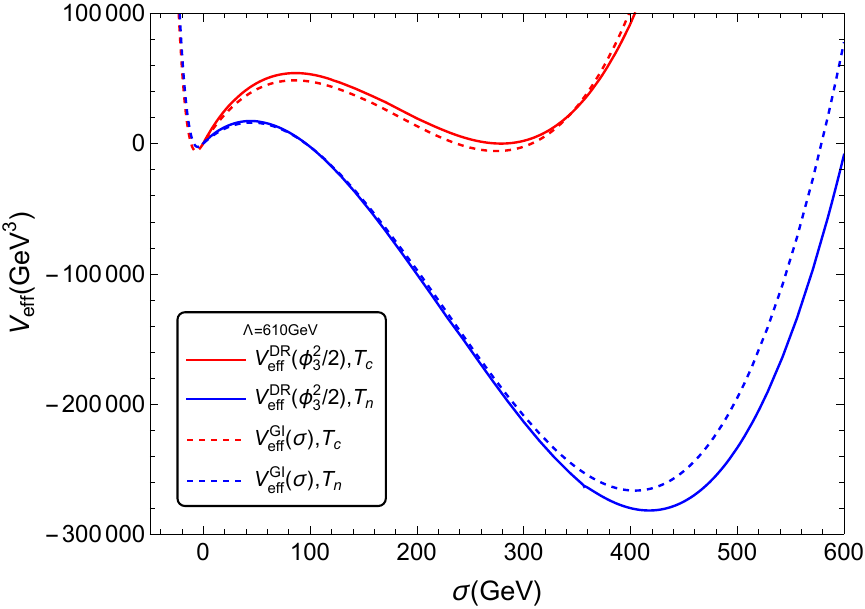}
    \includegraphics[width=0.4\linewidth]{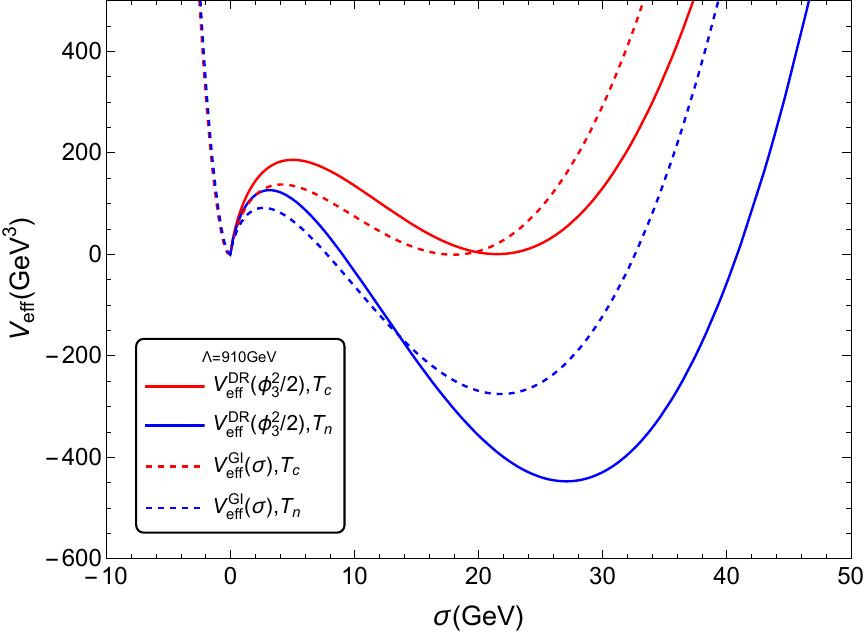}   
    \caption{The effective potential at $\Lambda=610$GeV and $910$GeV. The $V_{eff}^{DR}$ is 2-loop DR effective potential at light scale \cite{Qin:2024idc}, while $V_{eff}^{GI}$ is the GI potential given by Eq.~\eqref{veffg}.}
    \label{figveff2}
\end{figure}

In this appendix, we show the effective potential with $\Lambda=610$ GeV and $910$ GeV at temperature $T_c$ and $T_n$. In the Fig.~\ref{figveff2}, we plot the effective potential at temperature $T_c$ and $T_n$ with $\Lambda=610$ GeV and $910$ GeV. The GI potential of Fig.~\ref{figveff2} is defined in Eq.~\eqref{veffg}, and the relation $\sigma=\phi^2/2$ has been adopted to compare these results. And the potential $V_{eff}$ is obtained by the 2-loop order DR approach. In Fig.\ref{figveff2}, the two potentials with $\Lambda=610$ GeV in the broken phase are very similar. However, the GI potential in the symmetric phase shows a strong increase at small values of $\sigma$. The difference between these two potentials with  $\Lambda=910$ GeV is more significant than that of $\Lambda=610$ GeV, but this difference has less effect on the parameter $v/T$ in the Fig.~\ref{figtvbt}. 

If the equation of tree-level potential $dV_{tree}/d\phi=0$ has a non-trivial solution, the vev can be expanded by $\hbar-$expansion. At the 2-loop level, it becomes~\cite{Farakos:1994xh} 
\begin{equation}
\begin{aligned}
   V_{eff}(v^2)&=V_{tree}(v^2)+\hbar V_{1loop}(v^2)+\hbar^2 V_{2loop}(v^2)+\mathcal{O}(\hbar^3),\\
  v^2&=v_0^2+\hbar v_1^2+\hbar^2 v_2^2+\mathcal{O}(\hbar^3)\;.
\end{aligned}
\end{equation}
Then the condensate $\langle\Phi^\dag\Phi\rangle$ can be calculated at minima of the effective potential
\begin{equation}
   \langle\Phi^\dag\Phi\rangle=\frac{d V_{eff}(v^2)}{d\overline{\mu}_{h,3}^2}\; .
\end{equation}
The condensate $\langle\Phi^\dag\Phi\rangle$ is gauge invariant since the $m_\chi=0$ at this point.
However, this approach can only obtain physics at the minima of effective potential rather than the whole PT process including bubble nucleation.

\section{Nucleation rate}

\begin{figure}[!htp]
    \centering
    \includegraphics[width=0.4\linewidth]{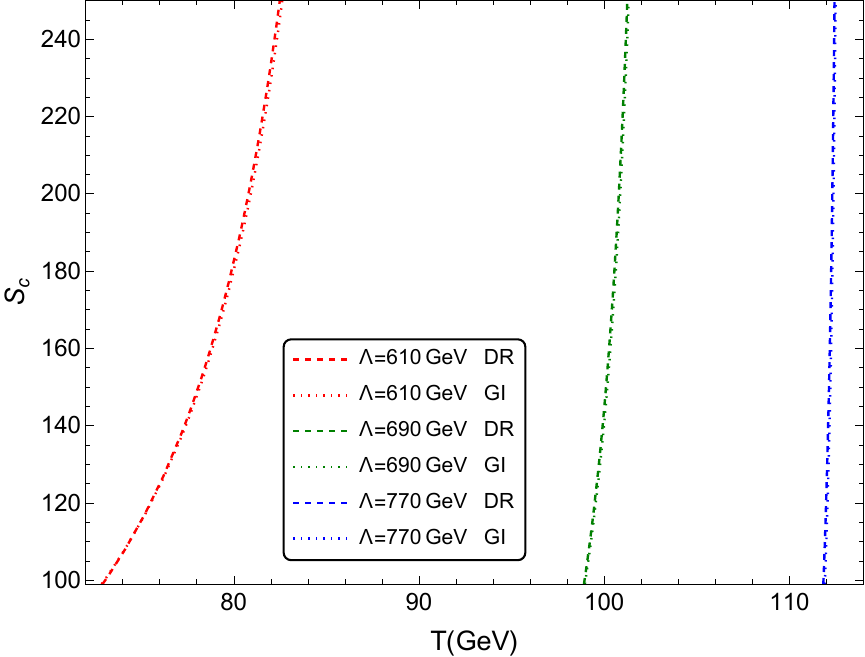}
    \includegraphics[width=0.4\linewidth]{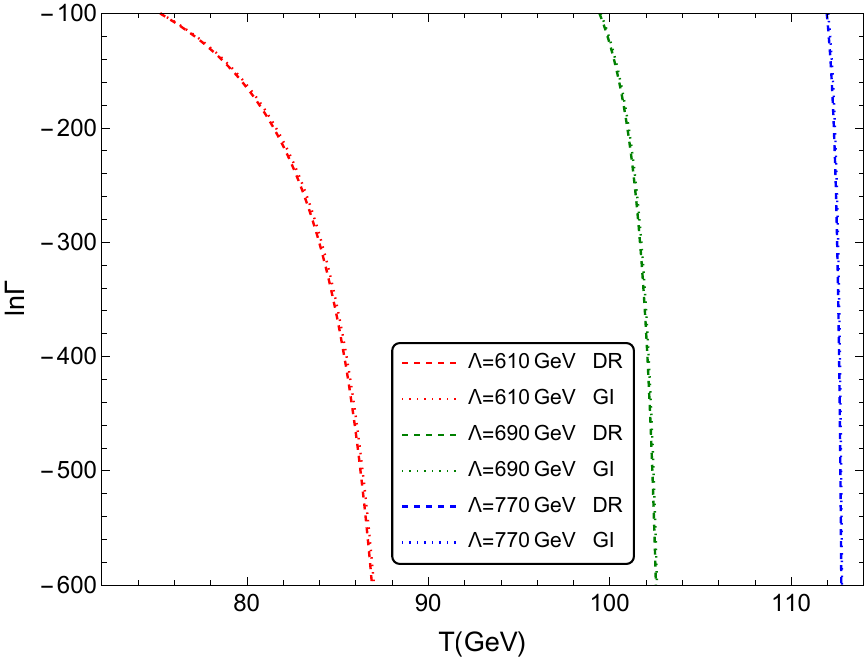}
    \caption{The action $S_c$ and nucleation rate $\Gamma$ as function of temperature $T$ with $\Lambda=610,690,770$GeV. The DR denotes the result of dimensional reduction and GI denotes the result of the gauge invariant approach.}
    \label{figscrate}
\end{figure}

We compared the difference between the DR and gauge invariant approach at the action $S_c$ and nucleation rate $\Gamma$ in this part. The action is calculated at the 2-loop order dimensional reduction with light scale, so it is a dimensionless quantity.  In the figure.\ref{figscrate}, we find the difference of $S_c$ and $\Gamma$ between the DR and GI is very slight and difficult to discern. This figure shows that as the temperature increases, the action increases while the nucleation rate decreases. The left one shows the slope of the curve increases with $\Lambda$, which makes $\beta/H$ increase with $\Lambda$, as shown in figure.\ref{alphabetah}.  The right one shows the nucleation rate decreases with increasing $\Lambda$. But the nucleation rate increases with $\Lambda$, even though the nucleation temperature $T_n$ also increases with $\Lambda$ in the figure.\ref{figrateA}.

\section{Electroweak sphaleron}
The sphaleron energy with the $U(1)_Y$ contribution reads,
\begin{equation}\label{spe3}
\begin{aligned}
E_{sph}^{EW}(\nu,T)=&\frac{4\pi v}{g}\frac{v(T)}{v}\int_{0}^{\infty}d\xi \left\{\left(\frac{8}{3}f^{\prime 2}+\frac{4}{3}f^{\prime 2}_3\right)
+\frac{8}{\xi^2}\left[\frac{2}{3}f_3^2(1-f)^2+\frac{1}{3}(f(1-f)+f-f_3)^2\right. \right]\\
&+\frac{4}{3}\left(\frac{g}{g^\prime}\right)^2\left[f_0^{\prime 2}+\frac{2}{\xi^2}(1-f_0)^2\right]+\frac{1}{2}\xi^2h^{\prime 2}+h^2\left[\frac{1}{3}(f_0-f_3)^2+\frac{2}{3}(1-f)^2\right]+\left.\frac{\xi^2}{g^2 v(T)^4}V(h,T)\right\}\;,\\
\end{aligned}
\end{equation}
where $\xi$ is a dimensionless parameter with $\xi=g v r $.
The $h,f,f_0,f_3$ are functions that enter the Higgs and gauge field configurations and can be obtained by solving
\begin{equation}\label{dipolefuns}
\begin{aligned}
&f^{\prime \prime}+\frac{1-f}{4\xi^2}\left[8(f(f-2)+f_3+f_3^2)+\xi^2h^2\right]=0 ,\\
&f^{\prime \prime}_3-\frac{2}{\xi^2}\left[3f_3+f(f-2)(1+2f_3)\right]-\frac{h^2}{4}(f_3-f_0)=0 ,\\
&f^{\prime \prime}_0+\frac{g^{\prime 2}}{4g^2} h^2(f_3-f_0)+2\frac{1-f_0}{\xi^2}=0 ,\\
&h^{\prime \prime}+\frac{2}{\xi}h^\prime-\frac{2}{3\xi^2}h\left[2(f-1)^2+(f_3-f_0)^2\right]-\frac{1}{g^2v(T)^4}\frac{\partial V(h,T)}{\partial h}=0\;,
\end{aligned}
\end{equation}
with boundary condition
\begin{equation}
\begin{aligned}
f(\xi)=0, h(\xi)=0, f_3(\xi)=0, f_0(\xi)=1&,\quad \xi\rightarrow 0\;,\\
f(\xi)=1, h(\xi)=1, f_3(\xi)=1, f_0(\xi)=1&,\quad \xi\rightarrow \infty\;.\\
\end{aligned}
\end{equation}

 \begin{figure}[!htp]
    \centering
    \includegraphics[width=0.4\linewidth]{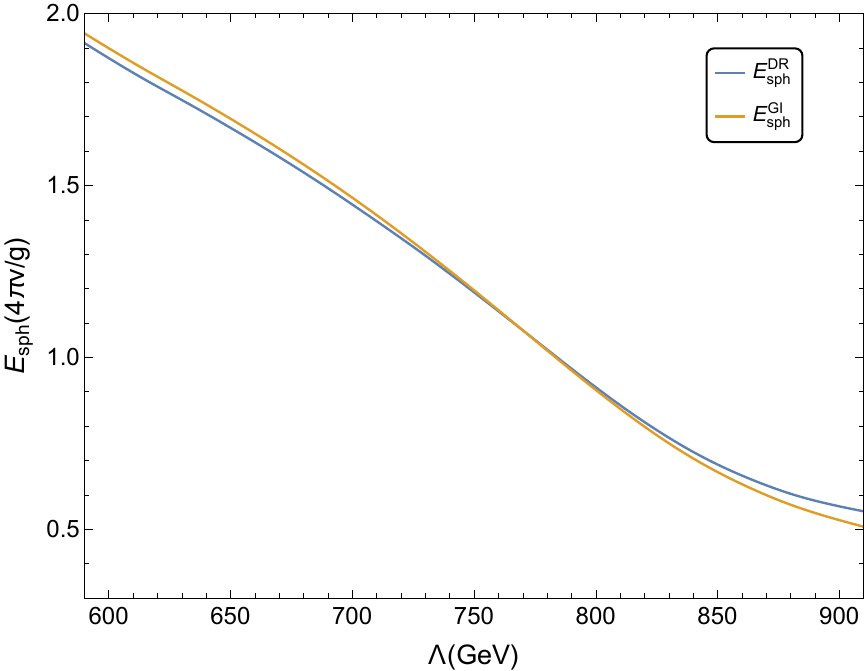}
      \includegraphics[width=0.4\linewidth]{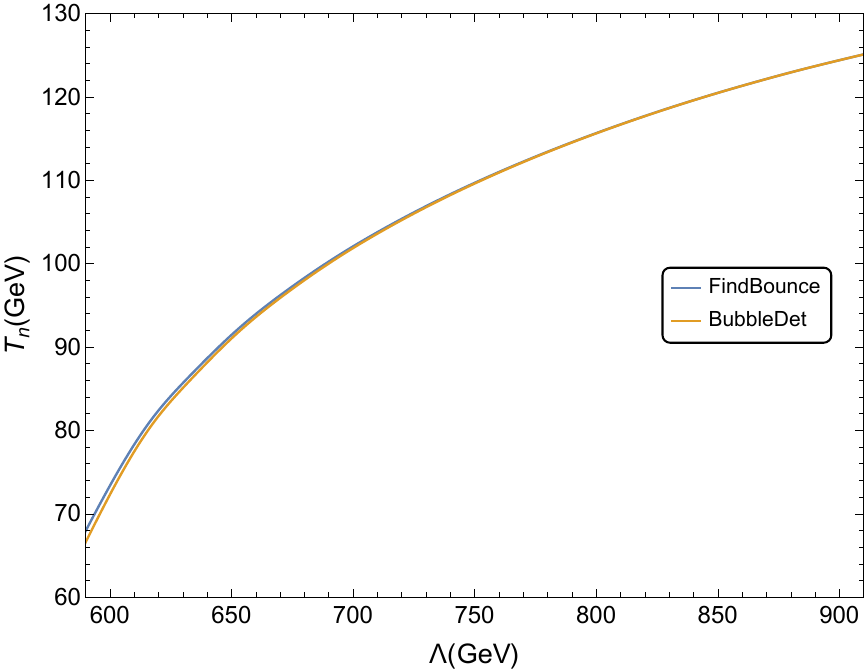}
    \caption{Left: The $E_{sph}$ as function of $\Lambda$ calculated with ``BubbleDet" with DR's and GI's potentials; Right: The $T_n$ as a function of NP scale with codes of ``BubbleDet" and ``FindBounce".}
    \label{figsp}
\end{figure}

The left plot of the Fig.\ref{figsp} shows the sphaleron energy $E_{sph}$ decreases with $\Lambda$ increase, and the GI $E_{sph}^{GI}$ is larger than it in DR $E_{sph}^{DR}$ at small $\Lambda$ region. With the $\Lambda$ increase, the $E_{sph}^{GI}$ is decrease faster than $E_{sph}^{DR}$, and it less than $E_{sph}^{DR}$ when $\Lambda\gtrsim 760$ GeV.  But this difference of sphaleron energy does not affect the limitation of the model parameter in Fig.~\ref{figpt}.
In the right panel of the Figure, we present the 
$T_n$ at different NP scales when we calculate the sphaleron energy. As a cross-check, we present both the results obtained with the code `FindBounce'' based on Polygonal bounce method~\cite{Guada:2020xnz} and the code ``BubbleDet'' based on Gelfand-Yaglom theorem~\cite{Ekstedt:2023sqc}.
The plot shows that there is a slight difference between the two codes for the calculation of the nucleation temperature $T_n$ in the small $\Lambda$ regions. However, this difference decreases with increasing $\Lambda$ until it is indistinguishable.

\section{Gravitational wave}
We consider three sources bubble collision, sound wave and turbulence for the production of GWs ~\cite{Wang:2020jrd,Bian:2021dmp,Zhou:2022mlz}.
The bubble walls collision term $\Omega_{co}$ is given by\cite{Huber:2008hg,Caprini:2015zlo,Kamionkowski:1993fg}
\begin{equation}\label{co}
\Omega h^2_{co}(f)\simeq 1.67\times 10^{-5}\left(\frac{\beta}{H}\right)^{-2}\left(\frac{\kappa_\phi\alpha}{1+\alpha}\right)^2\left(\frac{100}{g^*}\right)^{1/3}\frac{0.11v_w^3}{0.42+v_w^2}\frac{3.8(f/f_{co})^{2.8}}{1+2.8(f/f_{co})^{3.8}}\;,
\end{equation}
where the bubble wall velocity is defined as \cite{Lewicki:2021pgr}
\begin{equation}\label{vw}
v_w=\begin{cases}
    \sqrt{\frac{\Delta V}{\alpha \rho_r}}&, \quad \sqrt{\frac{\Delta V}{\alpha \rho_r}}<v_J(\alpha)\\
    1&,\quad \sqrt{\frac{\Delta V}{\alpha \rho_r}}\geq v_J(\alpha)
\end{cases}    
\end{equation}
with the $\Delta V$ being the difference between the broken phase and symmetric phase and the Jouguet velocity $v_J(\alpha)$ is \cite{Lewicki:2021pgr,Steinhardt:1981ct}
\begin{equation}
v_J=\frac{1}{\sqrt{3}}\frac{1+\sqrt{3\alpha^2+2\alpha}}{1+\alpha}\;.
\end{equation}
 The peak frequency $f_{co}$ locates at
\begin{equation}
f_{co}=1.65\times 10^{-5} \text{Hz} \frac{\beta}{H}\left(\frac{0.62}{1.8-0.1v_w+v_w^2}\right)\left(\frac{T}{100GeV}\right)\left(\frac{g^*}{100}\right)^{1/6}\;.
\end{equation}
The sound wave contribution is given by
\begin{equation}\label{sw}
\Omega h_{sw}^2(f)=2.65\times 10^{-6}(H \tau_{sw})\left(\frac{\beta}{H}\right)^{-1}v_w\left(\frac{\kappa_\nu \alpha}{1+\alpha}\right)^2 \left(\frac{g^*}{100}\right)^{-1/3}\left(\frac{f}{f_{sw}}\right)^3\left(\frac{7}{4+3(f/f_{sw})^2}\right)^{7/2}\;,
\end{equation}
with the peak frequency being \cite{Hindmarsh:2013xza,Hindmarsh:2015qta,Hindmarsh:2017gnf}
\begin{equation}
 f_{sw}=1.9\times 10^{-5}\frac{\beta}{H}\frac{1}{v_w}\frac{T}{100}\left(\frac{g^*}{100}\right)^{1/6}\text{Hz}\;.   
\end{equation}
The $\tau_{sw}$ shows the duration of the sound wave from the phase transition, which is calculated as
\begin{equation}
\tau_{sw}=\text{min}\left[\frac{1}{H},\frac{R_*}{\overline{U}_f}\right]\;,
\end{equation}
where $H_*R_*=v_w(8\pi)^{1/3}(\beta/H)^{-1}$, and $\overline{U}_f$ can be approximated as \cite{Hindmarsh:2017gnf,Caprini:2019egz,Ellis:2019oqb}
\begin{equation}   
\overline{U}_f^2\approx\frac{3}{4}\frac{\kappa_\nu \alpha}{1+\alpha}\;,
\end{equation}
the $\kappa_\nu$ is given by \cite{Espinosa:2010hh}
\begin{equation}
\kappa_\nu=\frac{\sqrt{\alpha}}{0.135+\sqrt{0.98+\alpha}}\;.
\end{equation}
The magnetic hydrodynamic turbulence term is given by
\begin{equation}\label{turb}
\Omega h_{turb}^2(f)=3.35\times 10^{-4}\left(\frac{\beta}{H}\right)^{-1}\left(\frac{\epsilon\kappa_\nu\alpha}{1+\alpha}\right)^{3/2}\left(\frac{g^*}{100}\right)^{-1/3}v_w\frac{(f/f_{turb})^3(1+f/f_{turb})^{-11/3}}{1+8\pi f a_0/(a_* H_*)}\;,
\end{equation}
with the peak frequency \cite{Caprini:2009yp}
\begin{equation}
f_{tur}=2.7\times 10^{-5}\frac{\beta}{h_*}\frac{1}{v_w}\frac{T_*}{100}\left(\frac{g_*}{100}\right)^{1/6}\text{Hz}\;.
\end{equation}
The efficiency factor $\epsilon\approx 0.1$, redshift of the frequency is obtained as
\begin{equation}
h_*=(1.65\times 10^{-5}\text{Hz})\left(\frac{T_*}{100\text{GeV}}\right)\left(\frac{g_*}{100}\right)^{1/6}\;.
\end{equation}
The predicted GW spectrum can then be readily calculated from the three sources
discussed above \eqref{co},\eqref{sw} and \eqref{turb}, leading to
\begin{equation}\label{gw}
\Omega_{GW}h^2=\Omega h^2_{co}(f)+\Omega h_{sw}^2(f)+\Omega h_{turb}^2(f)\;.
\end{equation}

\end{document}